# Title

Infrared anomalies in ultrathin $Ti_3C_2T_x$ MXene films

# Authors

Meng LI[1], Tao CHENG[2], Gongze LIU[1], He HUANG[3], Keqiao LI[1], Yang LI[4,5] *, Jiayue YANG[2] * and Baoling HUANG[1, 6, 7,8] *

# Affiliations

[1]Department of Mechanical and Aerospace Engineering, The Hong Kong University of Science and Technology, Clear Water Bay, Kowloon, Hong Kong 999077, China

[2]School of Energy and Power Engineering, Shandong University, Jinan 250061, China

[3]School of Materials Science and Engineering, Beijing Advanced Innovation Center for Materials Genome Engineering, University of Science and Technology Beijing, Beijing 100083, China

[4]State Key Laboratory of Fluid Power and Mechatronic Systems, School of Mechanical Engineering, Zhejiang University, Hangzhou 310027, China

[5]Key Laboratory of Advanced Manufacturing Technology of Zhejiang Province, School of Mechanical Engineering, Zhejiang University, Hangzhou 310027, China

[6]HKUST Shenzhen-Hong Kong Collaborative Innovation Research Institute, Futian, Shenzhen 518055, China

[7]HKUST Foshan Research Institute for Smart Manufacturing, Hong Kong University of Science and Technology, Clear Water Bay, Kowloon, Hong Kong 999077, China

[8]Thrust of Sustainable Energy and Environment, Hong Kong University of Science and Technology (Guangzhou), Guangzhou, China

*Corresponding author. E-mail: meyangli@zju.edu.cn; jy_yang@sdu.edu.cn; mebhuang@ust.hk
**Abstract**

Visible transparent but infrared reflective materials are ideal candidates for both transparent conductive films and low-emissivity glass, which are highly desired in a broad variety of areas such as touchscreens and displays, photovoltaics, smart windows, and antistatic coatings. Ultrathin $Ti_3C_2T_x$ MXene films are emerging as promising low-emissivity transparent candidates. However, the fundamental IR properties of $Ti_3C_2T_x$ has not been revealed experimentally due to daunting challenges in the preparation of continuous, large-area, and ultrathin films of optical quality on flat substrates. Herein, we proposed a tape-free transfer method that can help prepare centimeter-size and ultrathin (down to 8 nm) $Ti_3C_2T_x$ films on diverse optical substrates. Benefitting from this method, the refractive index and permittivity for $Ti_3C_2T_x$ were successfully measured. $Ti_3C_2T_x$ films exhibit large in-plane permittivity in the IR region, yielding maximum IR reflectance of 88% for bulk films. Interestingly, three anomalies were found in ultrathin $Ti_3C_2T_x$ films: strong dispersion in the permittivity, interlayer space-dependent optical properties, and abnormally high IR absorption for a 15-nm-thick film. These anomalies are important guidelines in the design of $Ti_3C_2T_x$-based low-emissivity transparent films and other related devices, and may inspire other intriguing applications such as ultrathin IR absorption coatings and tunable IR optical devices.

**Introduction**

MXenes, a family of emerging two-dimensional (2D) materials[1], have attracted wide attention because of their high electrical conductivity, strong photo-thermal effects, high electromagnetic wave absorption, great stability, and facile synthesis process[2, 3]. Since Naguib et al.[4] synthesized the first type of MXene, over 30 types of MXenes with various compositions have been synthesized, as well as many forecasted MXenes in theory[1, 5, 6]. Among them, titanium carbide ($Ti_3C_2T_x$) is the most widely investigated MXene, which has delivered great breakthroughs for various fields, such as energy storage[7, 8, 9], electromagnetic interference (EMI) shielding[10, 11, 12], wearable electronics[13], and smart monitoring[14, 15].

Recently, the optical properties of $Ti_3C_2T_x$ films stacked from 2D nanosheets were also investigated. They exhibit high light absorption over the UV, visible, and part of near-IR regions, indicating excellent solar energy harvesting ability. Interestingly, further optical investigations in the mid-IR region (>2.5 μm) have shown that these $Ti_3C_2T_x$ assemblies have limited absorption but high reflection to mid-IR light. Referring to the Kirchhoff's law, a low mid-IR absorption leads to a low IR emissivity. As a result, a 15-μm-thick $Ti_3C_2T_x$ assembly shows a solar absorptance as high as 90%, but a low absorptance/emittance of 10%[16]. This prominent spectral selectivity of $Ti_3C_2T_x$ yields a much higher solar-harvesting efficiency compared to traditional blackbody-like solar absorbers such as carbon-based materials. First-principles calculations indicate that this sharp change in the absorption properties of thick $Ti_3C_2T_x$ films in the mid-IR region is attributed to the transition in the real part of permittivity to negative values, as well as the large imaginary part. Such an intrinsically selective material has aroused wide attention in other areas such as multi-spectral camouflage and thermal management[17, 18], and transparent conductive films[19, 20]. However, the theoretical calculations cannot accurately reveal the optical properties of $Ti_3C_2T_x$ because both perfect stoichiometry and constant interlayer space are assumed. In fact, $Ti_3C_2T_x$ is a non-stoichiometric 2D material with substantial Ti vacancies and tunable interlayers. The experimental measurement of wavelength-dependent optical properties (i.e., complex permittivity and refractive index) of $Ti_3C_2T_x$ films over the broad spectrum from UV and visible to IR regions is an essential prerequisite for the well design of $Ti_3C_2T_x$-based optical devices.

Continuous $Ti_3C_2T_x$ films in centimeter size of optical quality on smooth substrates (i.e., glass, quartz, and silicon wafer) are required for the accurate characterization of IR optical properties, as the diameter of polarized light beam in the IR ellipsometry is approximately several millimeters. However, so far, it is a non-trivial task to prepare such $Ti_3C_2T_x$ films. When the thickness reduces to around 10 nm, this task becomes more technically challenging as $Ti_3C_2T_x$ monolayers generally have both a lateral size of less than 10 μm and an irregular shape. Moreover, the preparation of such large-area and ultrathin $Ti_3C_2T_x$ film of optical quality can benefit not only the IR measurements but also the real applications in many areas such as transparent conductive films and low-E smart windows. Some methods have been reported to prepare $Ti_3C_2T_x$ films on glass, such as spin coating[21], , spray coating[22], and blade coating[23]. However, these methods suffer from low surface coverage, poor thickness control, or imperfect nanosheet alignment.

Herein, we developed a facile but effective transfer method to fabricate ultrathin $Ti_3C_2T_x$ films (as thin as 8.6 nm) on different substrates, such as glass, silicon (Si) wafers, and aluminum (Al) plates, with optical qualities, centimeter-scale size, and desired thickness. Enabled by this transfer method the optical properties of $Ti_3C_2T_x$ films in the mid-IR region were successfully measured. Interestingly, three anomalies were found in ultrathin $Ti_3C_2T_x$ films: large negative permittivity, interlayer space-dependent optical properties, and huge IR absorption for a 15-nm-thick film.

**Results**

As mentioned above, a large-area (centimeter-scale) continuous $Ti_3C_2T_x$ film of optical quality on smooth substrates (i.e., glass and silicon wafer) is required for the accurate characterization of mid-infrared (IR) optical properties using IR ellipsometry technique. However, when the thickness of $Ti_3C_2T_x$ films reduces to tens of nanometers, it is challenging to prepare them over a large area with perfect surface coverage and optical qualities on different substrates. This is because monolayer $Ti_3C_2T_x$ flakes generally have a highly irregular shape, small lateral size of less than 10 μm (Supplementary Fig. 2), but ultrasmall thickness of 1.4 nm (Supplementary Fig. 3, and ~2 nm by AFM in Supplementary Fig. 4). In this work, we proposed a tape-free but effective transfer method to produce large-area $Ti_3C_2T_x$ films of optical quality with a thickness as low as 8.6 nm on various substrates (Supplementary Fig. 1). First, the aqueous suspension of $Ti_3C_2T_x$ flakes was filtrated on mixed cellulose ester (MCE) membrane filters with pores' sizes of hundreds of nanometers under the vacuum assistance. After the removal of most water solvent through the filter channels, $Ti_3C_2T_x$ flakes were stacked layer-by-layer, forming a continuous and semi-dry thin film with the water uptake of around 300% under the environmental condition of 16 °C for ambient temperature and 90% for relative humidity (Supplementary Fig. 5; detailed process of experimental data can be discovered in Supplementary Materials). Then, a smooth pre-cleaned substrate with ideal flatness (i.e., the root-mean-square roughness of 0.47 nm for glass in Supplementary Fig. 6) was placed on the semi-dry $Ti_3C_2T_x$ film. When applying a light force (~0.3 MPa) on the backside of the substrate, $Ti_3C_2T_x$ films with desired thicknesses were completely transferred to the target substrate from the MCE filter as shown in Fig. 1b. After the transfer process, remaining water solvent inside $Ti_3C_2T_x$ films would totally evaporate on a hot plate at the temperature of 40-50 °C. As a result, $Ti_3C_2T_x$ films could be tightly attached to various smooth substrates for further measurements. X-Ray photoelectron spectroscopy (XPS) surface analysis of dry $Ti_3C_2T_x$ films demonstrates the atomic concentration of Ti (28.14%), C (39.53%), O (19.82%), F (8.63%), and Cl (3.89%) induced during the etching procedure (Supplementary Fig. 7).

To figure out the underlying mechanism of this tape-free transfer method, we observed the surface morphology of the $Ti_3C_2T_x$ film. As shown in Fig. 1c, there are widely distributed crumples on the surface of $Ti_3C_2T_x$ films (568-nm-thick) induced by large vacuum pressure. After the placement of the smooth substrate, semi-closed air gaps were confined between crumpled $Ti_3C_2T_x$ films and the substrate. When the substrate was pressed, air was expelled from the gap, forming negative pressure between the $Ti_3C_2T_x$ film and the substrate (Fig. 1a). As the two interfaces approach to each other, strong Van deer Waals force was built between them. Moreover, lower surface tension of water was generated when a substrate was attached to semi-dry $Ti_3C_2T_x$ films, especially for hydrophilic substrates such as glass (Supplementary Fig. 8). As a result, $Ti_3C_2T_x$ films were successfully transferred due to the collective power of negative pressure, Van deer Waals force, as well as lower surface tension. This transfer method is also compatible with other smooth substrates such as wafers and Al plates with various shapes (Supplementary Fig. 9).

As shown in Fig. 1c, the $Ti_3C_2T_x$ film after transfer has a smoother surface than the film before transfer, contributing to lower infrared emittance compared to crumpled surface of $Ti_3C_2T_x$ films[16], which is favorable in low-E applications such as low-E glasses. Fig. 1e compares the cross-sectional morphology of the $Ti_3C_2T_x$ film before transfer with that after transfer on a wafer. The result verifies that the transferred $Ti_3C_2T_x$ film adheres tightly to the substrate, and maintains excellent alignment and thickness almost unchanged along the cross-section. A few air voids inside the $Ti_3C_2T_x$ films are caused by the Van der Waals force between the MCE membrane and $Ti_3C_2T_x$ film when the two separates from each other.

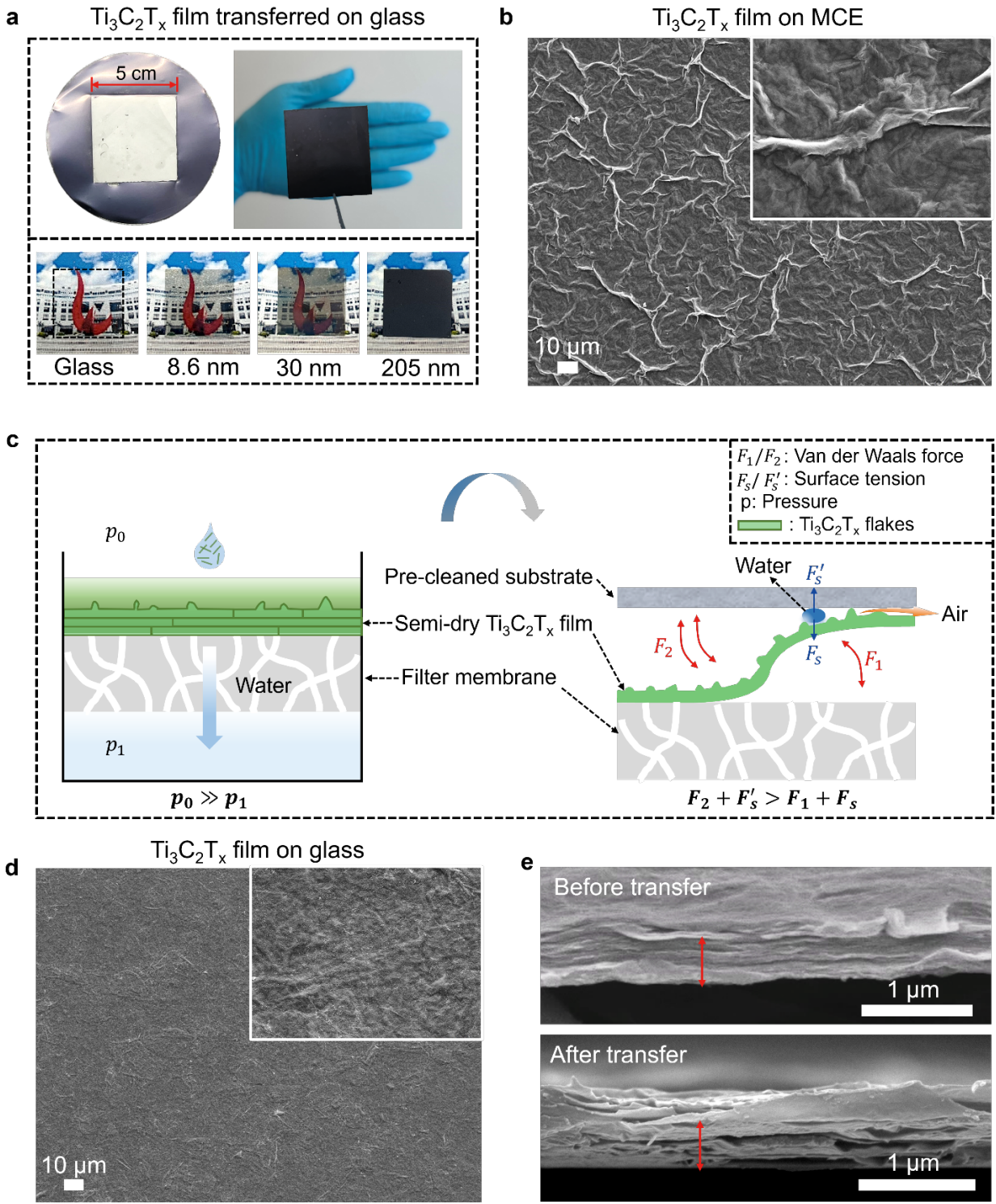

**Fig. 1.** Preparation of $Ti_3C_2T_x$ films. (**a**) Fabrication mechanism of the transfer method. (**b**) Images of transferred $Ti_3C_2T_x$ films on glass. (**c**) SEM images of $Ti_3C_2T_x$ films on porous membranes (before transfer). (**d**) SEM images of $Ti_3C_2T_x$ films on the wafer (after transfer). (**e**) Cross-sectional

SEM image of $Ti_3C_2T_x$ films on porous membranes (before transfer) and on the wafer (after transfer), and the thickness of the $Ti_3C_2T_x$ film before transfer is averagely ~568 nm.

Complex permittivity ($\tilde{\varepsilon} = \varepsilon_1(\omega) + i\varepsilon_2(\omega)$) or refractive index ($\tilde{n}(\omega) = n(\omega) + ik(\omega)$) over the wide wavelength range varying from visible (0.37-0.76 μm) to near-infrared (NIR, 0.76-2.5 μm) and mid-infrared (MIR, 2.5-20 μm) regions is the most fundamental parameter for optical materials like $Ti_3C_2T_x$. Fig. 2a shows the mechanism of permittivity measurement performed in an ellipsometric spectrometry using the non-destructive optical technique. Firstly, a beam of polarized light from the polarizer (incident light) illuminates on the surface of a $Ti_3C_2T_x$ film. The reflected part of incident light by the $Ti_3C_2T_x$ film is collected by a rotating analyzer. The terminal module of the detector can identify the polarization state change between the incident light and reflected light by our samples, known as the amplitude of $\psi$, and phase of $\Delta$[24]. In the ellipsometry, $\psi$ and $\Delta$ of certain materials are extracted as a function of wavelength in the material system to form an ideal curve[25] (Supplementary Note 2). The experimental data collected during the measurement can be regulated to fit the ideal curve by tuning either the thickness or optical constants of our $Ti_3C_2T_x$ films. Since $Ti_3C_2T_x$ films are too flexible to be free-standing and the influence of membrane filters on optical measurement of $Ti_3C_2T_x$ films could not be eliminated, they were transferred to normalized glass substrates integrally. Therefore, to obtain the best approximation of the intrinsic permittivity ($\tilde{\varepsilon}$) of $Ti_3C_2T_x$ films, the complex permittivity $\tilde{\varepsilon}$ of glass was constructed using a multi-oscillator model first. Afterwards, $\tilde{\varepsilon}$ of $Ti_3C_2T_x$ films was described via B-splines, which can effectively describe complex curves in absorption spectra while simultaneously preserving Kramers-Kronig consistency[26].

According to the molecular structure of $Ti_3C_2T_x$ in Fig. 2b, when the incident light is along the z direction, the electric field orients in the x-y plane. Namely, the optical characteristics of reflectance (R)/transmittance (T)/absorptance (A) are determined by the in-plane dielectric permittivity. For the first time, the in-plane broadband permittivity of stacked $Ti_3C_2T_x$ films was measured by visible/NIR ellipsometer and MIR ellipsometer in the wavelength range of 0.37-1.5 μm and 1.5-20 μm, respectively. Fig. 2c and d show the permittivity and refractive index of an optically thick (568 nm) $Ti_3C_2T_x$ film. Over the wide wavelength range from 0.37 to 20 μm, the real part of the permittivity shows a steady decline trend with a transition from positive values to negative values at around 1.1 μm, indicating a shift from dielectric (absorption) to metallic (reflection) feature. The imaginary part of complex permittivity constantly increases from zero to above 500. Overall, the permittivity profiles of $Ti_3C_2T_x$ are similar to those of metals such as gold and silver[27]. However, the transition wavelength that corresponds to plasma frequency of $Ti_3C_2T_x$ is longer than that of metals (around 200 nm for Au, and 330 nm for Ag) due to the relatively lower free electron density. Both the real part $n(\omega)$ and the imaginary part $k(\omega)$ of refractive index are quite small in the visible and NIR regions, but sharply increased in the MIR region. In addition, $k(\omega)$ is larger than $n(\omega)$ over the entire IR region due to the negative real part of complex permittivity.

The reflectance/transmittance/absorptance spectra of $Ti_3C_2T_x$ films on glass (Supplementary Fig. 10, the transmittance of glass is 92% without absorption.) in the wavelength range of 0.3-2.5 and 2.5-20 μm were collected by a UV-visible-NIR spectrometer and a FTIR spectrometer, respectively. According to Supplementary Note 3, in high-frequency region, such as UV, visible, and part of NIR regions for $Ti_3C_2T_x$, little light can be reflected by $Ti_3C_2T_x$ films as $n(\omega) \approx 1$, and $n(\omega) \gg k(\omega)$ ($R(\omega) = \frac{(n(\omega)-1)^2+k^2(\omega)}{(n(\omega)+1)^2+k^2(\omega)}$), yielding high absorption of around 80% (Fig. 2e). In long-wavelength NIR region (1.1-2.5 μm), the absorption of $Ti_3C_2T_x$ films drops considerably to around 20% with the increase of reflectance. The increase in the reflection is enhanced by the oscillation of free carriers[21], which is consistent with the transition wavelength of $\varepsilon_1$ at around 1.1 μm. Different from the spectral response in 0.3-2.5 μm, the optical characterization of $Ti_3C_2T_x$ films in Fig. 2f suggests strong reflectance of around 88% due to negative and large real part of $\varepsilon_1(\omega)$ as well as $k^2(\omega) \gg 4n(\omega)$ over the IR region varying from 2.5 to 16.7 μm, along with the emittance as low as 12%. The optical properties of 568-nm-thick $Ti_3C_2T_x$ films were also calculated using $R(\omega) = \frac{(n(\omega)-1)^2+k^2(\omega)}{(n(\omega)+1)^2+k^2(\omega)}$ and they showed great agreement with experimental spectra. The spectral selectivity of high absorption of 80% in 0.3-1.1 μm and low emittance of 12% in 2.5-16.7 μm makes $Ti_3C_2T_x$ of great potential employed in the fields of solar harvesting[16], and thermal camouflage scenarios[28, 29].

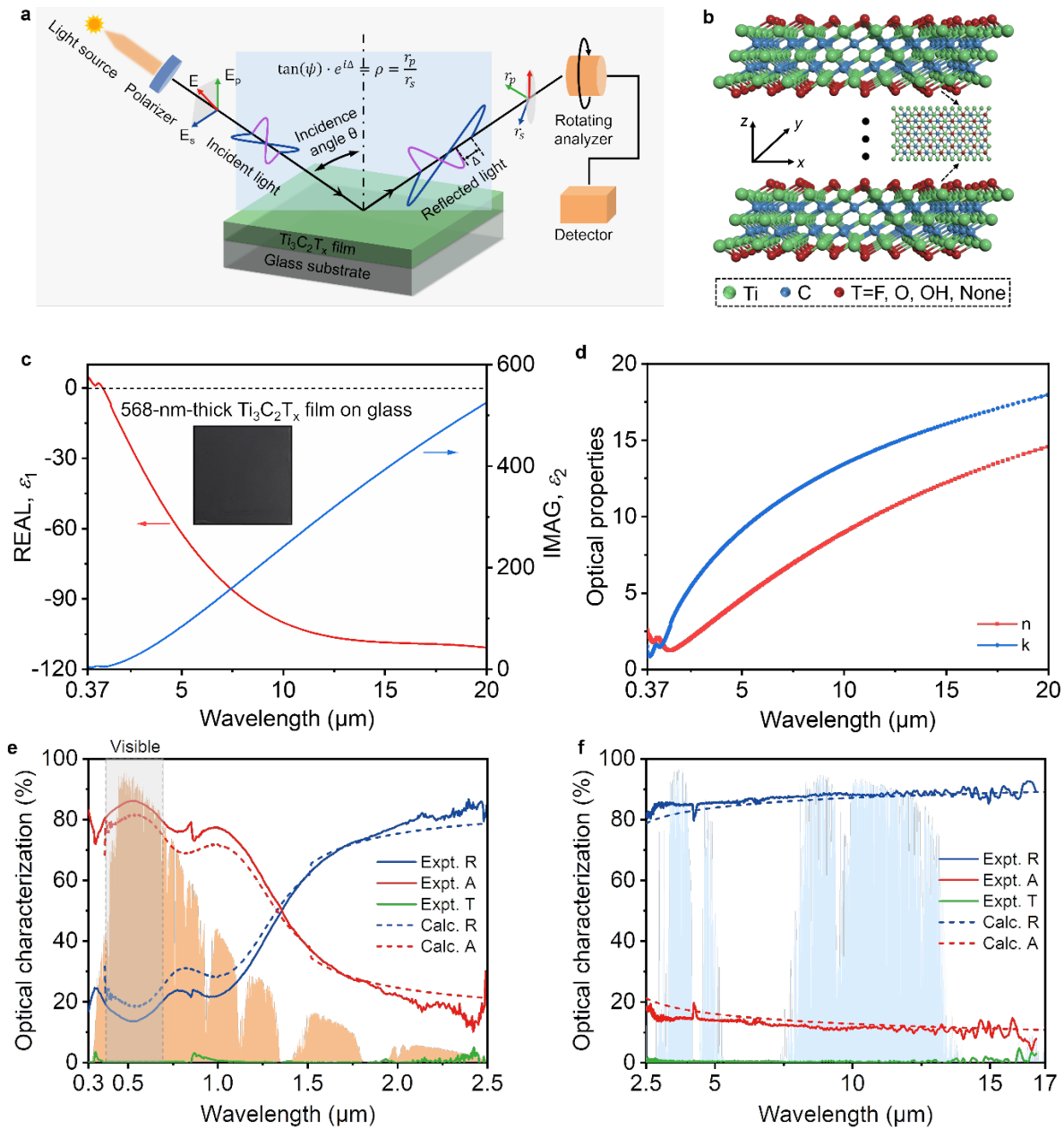

**Fig. 2.** Broadband optical properties of $Ti_3C_2T_x$ films in the UV-Vis-NIR and IR region. (**a**) Mechanism of optical properties measurement in the IR region. (**b**) Molecular structure of multi-layer $Ti_3C_2T_x$ films. (**c**) Refractive index of 568-nm-thick $Ti_3C_2T_x$ films in 0.3-20 µm. (**d**) Permittivity of 568-nm-thick $Ti_3C_2T_x$ films in 0.3-20 µm. (**e**) Experimental (abbreviated as Expt. in solid lines) and calculated (abbreviated as Calc. in dash lines) optical characterizations of 568-nm-thick $Ti_3C_2T_x$ films on the glass substrate in 0.3-2.5 µm. R, A, T indicate reflectance,

absorptance, and transmittance. (**f**) Experimental and calculated optical characterizations of 568-nm-thick $Ti_3C_2T_x$ films on porous MCE membranes in 2.5-20 μm.

This transfer method enables the preparation of large-area ultrathin $Ti_3C_2T_x$ films down to 8.6 nm made of a few layers on different substrates. The spectral response of ultrathin $Ti_3C_2T_x$ films over the broad band (0.3-20 μm) is of great importance for many applications such as low-E glass and transparent IR camouflage coating. However, the optical characterization measurement of ultrathin transparent films is technically challenging. For one thing, it is difficult to prepare free-standing ultrathin $Ti_3C_2T_x$ films for straightforward optical measurements. An alternative approach is to use optically transparent substrates. For another, there are few substrates that have high transparency (>90%) over the wide band from 0.3 to 20 μm. For example, the glass and quartz substrates are transparent at wavelengths below 2.7 μm but absorptive in the mid-IR region, while the zinc selenide (ZnSe) glass with antireflection coatings is highly transparent in the wavelength range of 2-13 μm. To address this challenge, we employed glass substrates for the measurements over the UV-visible-NIR range (0.3-2.5 μm), while mid-IR nonreflective substrates of polyvinylidene difluoride (PVDF) films with mid-IR reflectance of less than 2% (Supplementary Fig. 11), as well as nanoporous polyethylene (nanoPE) films with mid-IR transmittance of over 96% (Supplementary Fig. 12) for mid-IR measurements (2.5-20 μm).

Fig. 3a shows the solar transmission spectra of $Ti_3C_2T_x$ films with thickness varying from 8.6 to 1183 nm, as well as that of other ultrathin $Ti_3C_2T_x$ films thinner than 20 nm (Supplementary Fig. 13). The transmittance of ultrathin $Ti_3C_2T_x$ films such as 8.6 and 15.5 nm over the entire solar spectrum (0.3-2.5 μm) is as high as 87% and 65%, respectively. The solar transmittance of $Ti_3C_2T_x$ films decreases with the thickness and becomes totally opaque when the thickness is above 205 nm. The reflection spectra of $Ti_3C_2T_x$ films in the mid-IR region is measured on the nonreflecting PVDF substrates (Fig. 3b). It is notable that 62% of infrared light could be strongly reflected by a thin 28.8-nm-thick $Ti_3C_2T_x$ film. With the increase of film thickness to over 200 nm, the IR reflectance reaches a plateau of 88%, indicating a low IR emittance of 12%. For those ultrathin films with thickness below 20 nm, the IR reflectance is relatively low due to both high IR absorption and transmission. For example, the IR reflectance of 8.6 and 15.5-nm-thick $Ti_3C_2T_x$ films is approximately 24% and 28%, respectively. The infrared images of $Ti_3C_2T_x$ films of various thicknesses on glass substrates at a 100-°C hot plate were shown in Fig. 3d. Since the IR emittance of glass is as large as 86%, it exhibits a high apparent temperature of 95 °C. With the increased thickness of $Ti_3C_2T_x$ films, more and more infrared light is reflected to surroundings. Namely, the apparent temperature reduces with the increase of film thickness due to low infrared emittance of $Ti_3C_2T_x$, and tends to be steady at around 40 °C for $Ti_3C_2T_x$ films thicker than 200 nm, which is consistent with the results in Fig. 3b.

To verify the thickness dependence of IR reflection, we calculated the IR reflection spectra of $Ti_3C_2T_x$ films using the complex permittivity of a 568-nm-thick film. As expected in Supplementary Fig. 14, the calculated IR reflection increases with the film thickness. Although this thickness-dependence tendency agrees with the measured results, the calculated IR reflection of each film slightly differs from the measured data, and this deviation becomes more evident as the film thickness decreases (Fig. 3c). One hypothesis is that the intrinsic optical properties, *i.e.,* dielectric permittivity $\tilde{\varepsilon}$ or refractive index $\tilde{n}$, of $Ti_3C_2T_x$ films is also dependent on the film thickness. To validate this assumption, various $Ti_3C_2T_x$ films were transferred to glass substrates with optical quality (Supplementary Fig. 15, the root-mean-square roughness detected by atomic force microscopy ranges from 3.8 nm for an 8.6-nm-thick film to 10 nm for a 60-nm-thick film.), and their permittivity $\tilde{\varepsilon}$ was measured by an IR ellipsometer. As shown in Fig. 3e and f, their $\tilde{\varepsilon}$ is independent on the film thickness at wavelengths below 1.4 μm, but sensitive at longer wavelengths. For those thinner films, the absolute values of both the real and imaginary parts in the IR region are much larger than those for thicker films, leading to stronger metallic properties. For their complex refractive index derived from the permittivity in Supplementary Fig. 16, both the real and imaginary parts reduce with the increase of film thickness in the IR region and converges to that of optically thick $Ti_3C_2T_x$ films. As a result, thin $Ti_3C_2T_x$ films lower than 100 nm exhibit the ability of strong IR reflection, such as 61.6% for 28.8 nm and 71% for 50.4-nm-thick $Ti_3C_2T_x$ films. These films of large area and high optical quality result in high credibility of their refractive index data with small mean-squared error (MSE) as low as 0.72 (Supplementary Table. S1). Moreover, we re-calculated the reflectance of both an ultrathin (40 nm) and a thick (400 nm) $Ti_3C_2T_x$ film with the corresponding permittivity, validating that higher IR reflectance of $Ti_3C_2T_x$ films with the same thickness is partially attributed to larger dielectric permittivity (Supplementary Fig. 17).

To explore the underlying mechanism behind the permittivity variation of $Ti_3C_2T_x$ films mentioned above, the interlayer spaces of these films were analyzed from XRD patterns. As shown in Fig. 3g, the (002) diffraction peaks of $Ti_3C_2T_x$ films gradually increase with their thickness varying from 6.1° for 28.8 nm to 6.4° for 803 nm, indicating a decrease in the interlayer space from 14.4 Å for ultrathin $Ti_3C_2T_x$ films to 13.7 Å for thick $Ti_3C_2T_x$ films. This difference in the interlayer space may be induced during the fabrication process. Thick $Ti_3C_2T_x$ films on porous MCE membranes were filtrated under a higher vacuum pressure for a longer time to remove the solvent, resulting in denser stacking of $Ti_3C_2T_x$ flakes. According to the first-principles molecular dynamics model (Supplementary Note 2), smaller interlayer space for thick $Ti_3C_2T_x$ films yields lower total dipole moment, and ultimately contributes to smaller imaginary part of complex permittivity according to the formula in the classic limit as $\varepsilon_2(\omega) = \frac{2\pi\omega}{3Vk_BT} \int_{-\infty}^{\infty} dt e^{-i\omega t} \langle \mathbf{M}(t) \cdot \mathbf{M}(0) \rangle$[30]. Meanwhile, the real part $\varepsilon_1(\omega)$ of complex permittivity can be derived from $\varepsilon_2(\omega)$ based on the Kramers-Kronig (KK) dispersion relations as $\varepsilon_1(\omega) = \frac{2}{\pi} \int_0^{\infty} d\omega' \frac{\omega' \varepsilon_2(\omega')}{\omega'^2 - \omega^2} + \varepsilon_\infty$[31], from where it can be conducted that the real part of $\varepsilon_1(\omega)$ drops synchronously with $\varepsilon_2(\omega)$. As a result, the absolute values of both $\varepsilon_1(\omega)$ and $\varepsilon_2(\omega)$ decrease with the thickness of $Ti_3C_2T_x$ films due to their reduced interlayer space. To eliminate the influence from the mass (or thickness) of $Ti_3C_2T_x$ films, we fabricated $Ti_3C_2T_x$ films with the same mass of 2 mg for different filtration durations including 60, 80, and 100 seconds (Supplementary Fig. 18). As expected, with the increase of the filtration duration from 60 to 100 seconds, the interlayer space decreases from 15.5 to 14.6 Å, and therefore the permittivity decreases in general. This result reveals that the permittivity variation is caused by the difference in the interlayer space instead of the film thickness itself.

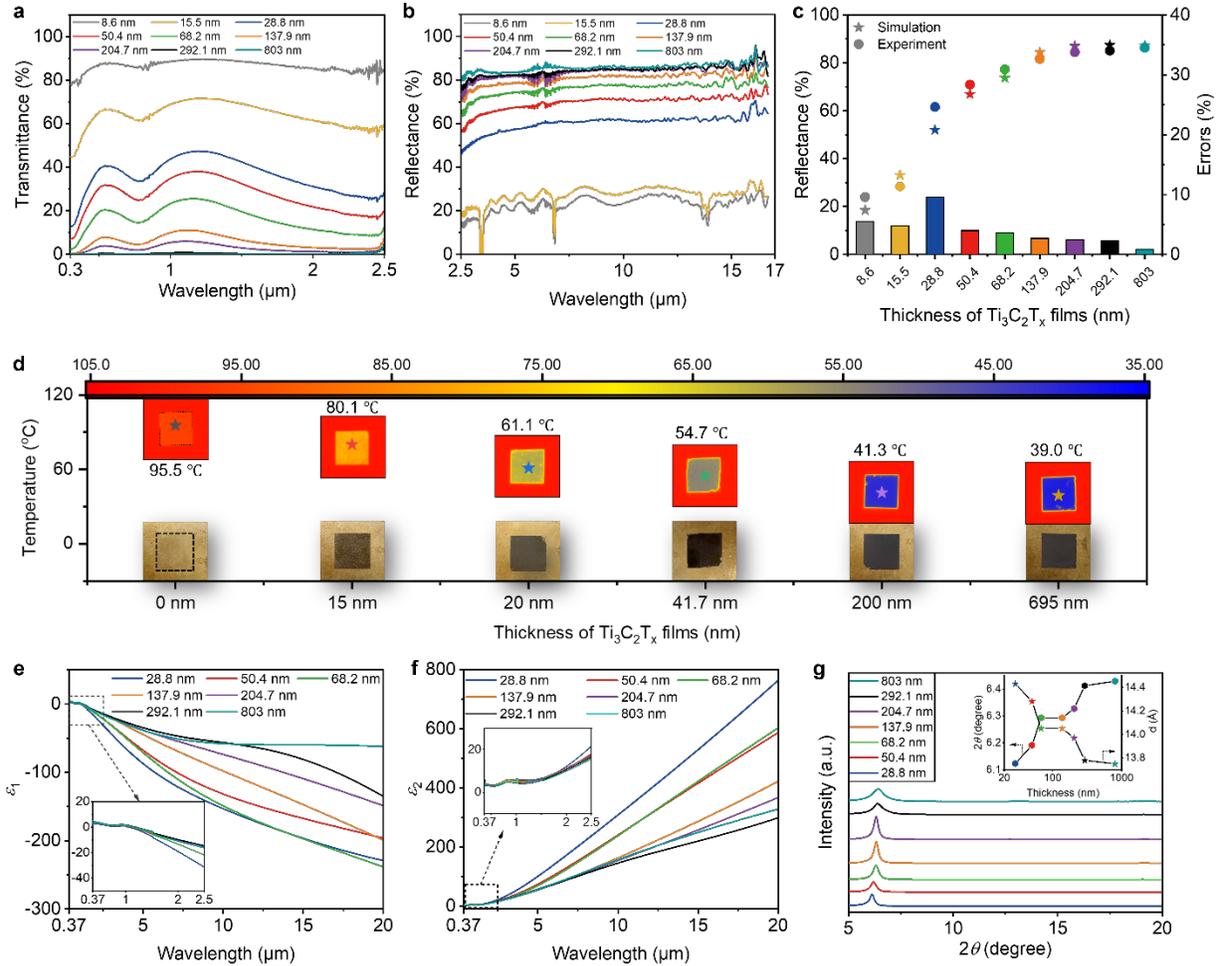

**Fig. 3.** Broadband optical characterizations' variation of $Ti_3C_2T_x$ films with different thickness. (**a**) Solar transmittance of $Ti_3C_2T_x$ films with various thickness. (**b**) Infrared reflectance of $Ti_3C_2T_x$ films with various thickness by experimental measurements. (**c**) Average IR reflectance of various $Ti_3C_2T_x$ films and their errors between experiment and simulation results. (**d**) Infrared images of the glass (0 nm) and $Ti_3C_2T_x$ films on glasses on the hot plate with the temperature of 100 °C. (**e**) Real part and (**f**) Imaginary part of $Ti_3C_2T_x$ films with various thickness. (**g**) XRD patterns of $Ti_3C_2T_x$ films with different thickness.

Besides strong reflective response to infrared light, low IR absorptance/emittance of ultrathin $Ti_3C_2T_x$ films is also critical for transparent low-E scenarios. Since ultrathin $Ti_3C_2T_x$ films are too fragile to be free-standing and partially transparent to infrared light, an IR-transparent supportive substrate of 12-μm-thick nanoPE films with an average transmittance of 96.6% in 2.5-20 μm was utilized to measure both the IR reflectance and transmittance of pristine ultrathin $Ti_3C_2T_x$ films.

Ti$_3$C$_2$T$_x$ films with different thickness on nanoPE films was prepared by vacuum-assist filtration method and their IR transmittance spectrum demonstrated that the IR transmittance of Ti$_3$C$_2$T$_x$ films decreases from 80.4% to 4.1% as the thickness increases from 5.0 to 68.2 nm (Supplementary Fig. 19). Fig. 4b shows the IR absorptance/emittance spectra of ultrathin Ti$_3$C$_2$T$_x$ films with different thicknesses. Strikingly, the absorptance/emittance first increases with the thickness of Ti$_3$C$_2$T$_x$ films until a maximum value of 37.5% for a 15-nm-thick Ti$_3$C$_2$T$_x$ film, and then gradually decreases. For optically thick Ti$_3$C$_2$T$_x$ films of hundreds of nanometers, their absorptance/emittance properties are independent on the film thickness, converging on a steady value of around 12% (see details in Supplementary Fig. 20).

One assumption for this "up and down" trend of IR absorptance for ultrathin Ti$_3$C$_2$T$_x$ films is that more electron collisions induced by higher carrier concentration in ultrathin Ti$_3$C$_2$T$_x$ films contribute to higher absorption of infrared photons. To adequately verify this hypothesis, both their carrier concentration and mobility were derived from the Hall effect measurement. As shown in Fig. 4c, a 17.7-nm-thick Ti$_3$C$_2$T$_x$ film exhibits a high carrier concentration of 2.08×10$^{28}$ m$^{-3}$, and the carrier concentration roughly drops with the increase of film thickness. By contrast, high carrier mobility was achieved in those thick films with low carrier concentrations, such as a large value of 26 cm$^2$ V$^{-1}$ s$^{-1}$ for a 1570-nm-thick Ti$_3$C$_2$T$_x$ film (Supplementary Fig. 21). Referring to Sommerfeld theory (Supplementary Note 4), the mean free path $l$ (m) of carriers in Ti$_3$C$_2$T$_x$ films is calculated as[32]:

$$\frac{\sigma_0}{l} = \frac{n\epsilon^2}{m\bar{v}} = \left(\frac{8\pi}{3}\right)^{\frac{1}{3}} \frac{\epsilon^2 n^{\frac{2}{3}}}{h} \tag{1}$$

where $m$ denotes the mass of one electron, $\bar{v}$ is the electron velocity at the surface of Fermi distribution, $h$ represents the Planck's constant, $\epsilon$ is the electron charge, and $n$ is the carrier concentration. As a result, $l$ in Ti$_3$C$_2$T$_x$ films increases with their thickness, and the $l$ of carriers in a 17.7-nm-thick Ti$_3$C$_2$T$_x$ film is only 0.25 nm, which is lower than the thickness of Ti$_3$C$_2$T$_x$ monolayers (~1 nm). With the increase of film thickness, the $l$ increases and reaches a plateau of around 5 nm in optically thick Ti$_3$C$_2$T$_x$ films of above 1000 nm. According to the theory of Matthiessen's rule[33], the effective $l$ of electrons ($\frac{1}{l_{eff}} = \frac{1}{l_{bulk}} + \frac{1}{l_{boundary}}$) in thin films is mainly determined by two parts, i.e., electron scattering ($l_{bulk}$) and scattering from the boundary of thin films ($l_{boundary}$). In thick Ti$_3$C$_2$T$_x$ films, moving electrons can travel freely before colliding with other electrons. Therefore, their $l_{eff}$ is dominated by the electron scattering in thick films, and is nearly equal to the $l_{bulk}$ = 5 nm in this work. However, the moving electrons in ultrathin $Ti_3C_2T_x$ films can be strongly scattered by the boundary of films owing to both ultrasmall thickness and high carrier concentrations as shown in Fig. 4d. As a result, the $l_{eff}$ in ultrathin $Ti_3C_2T_x$ films are quite short, such as 0.25 nm for a 17.7-nm-thick $Ti_3C_2T_x$ film. Since part of the electrons' kinetic energy is dissipated in the form of heat during the collisions with boundaries, i.e., inelastic collisions[34], more IR photons can be absorbed to excite those collided electrons, yielding higher IR absorptance for ultrathin $Ti_3C_2T_x$ films.

Apart from electronic properties, their sheet resistance was investigated by a four-probe method (inset, Fig. 4e). With the increase of film thickness from 15 nm to hundreds of nanometers, the sheet resistance per square area decreases from 182.9 to 1.5 Ω □$^{-1}$ (Supplementary Fig. 22), while the sheet conductance increases from the minimum value of 0.0055 S to 0.65 S. The average bulk conductivity is approximately 4962±220 S cm$^{-1}$ (Fig. 4e), which is comparable to previous research[35]. For a 15-nm-thick $Ti_3C_2T_x$ film, its sheet resistance of 182.9 Ω □$^{-1}$ accounts for approximately half of the free-space intrinsic impedance $\eta_0$ (377 Ω □$^{-1}$). This phenomenon further demonstrates the reasonability of its maximum IR absorptance of 37.5% according to the impedance matching model[36] in Supplementary Note 5. However, a maximum absorptance of 50% is derived from the impedance matching model[36] when the sheet resistance of $Ti_3C_2T_x$ films is comparable to half of the free-space intrinsic impedance $\eta_0$. The variations in IR absorptance between experimental and numerical calculation results for thin films with thickness below 50 nm can be explained by oxidation from $Ti_3C_2T_x$ to titanium dioxide ($TiO_2$) particles, especially around the boundary of $Ti_3C_2T_x$ flakes for thin films. We then numerically calculated the IR absorptance/emittance of various $Ti_3C_2T_x$ films varying from a monolayer (1 nm) to a bulk film of 1500 nm by finite-difference time-domain (FDTD) method using the refractive index of 568-nm-thick $Ti_3C_2T_x$ films (Supplementary Fig. 23). As expected, the average IR reflectance increases with the film thicknesses, while the average IR transmittance decreases (Fig. 4f). As a result, the average IR absorptance of $Ti_3C_2T_x$ films reaches a maximum of 47.6% at 15 nm thick, which is agree with the tendency of experimental results.

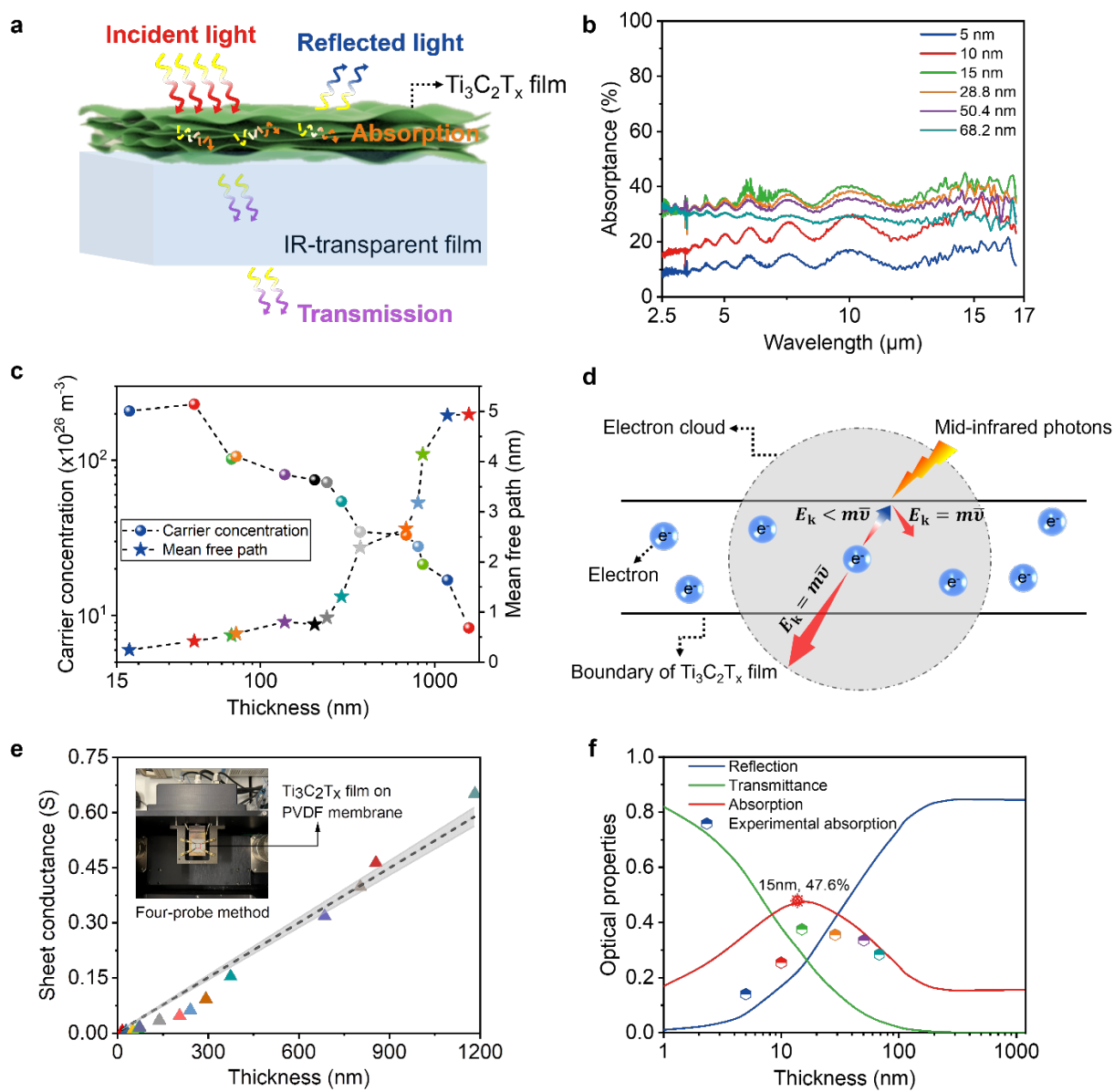

**Fig. 4**. Discipline of infrared absorptance/emittance of multi-layer $Ti_3C_2T_x$ films. (**a**) Schematic diagram of IR measurement for ultrathin $Ti_3C_2T_x$ films on nanoPE membranes. (**b**) Infrared absorptance/emittance of ultrathin $Ti_3C_2T_x$ films. (**c**) Carrier concentration and mean free path of $Ti_3C_2T_x$ films with different thickness. (**d**) Mechanism of electrons' movement in ultrathin $Ti_3C_2T_x$ films. (**e**) Sheet conductance of $Ti_3C_2T_x$ films varying with their thickness. (**f**) FDTD calculation and experimental results of optical characterizations for $Ti_3C_2T_x$ films with different thickness.

**Discussion**

We developed a transfer method with the assistance of vacuum filtration to fabricate large-size and optical-quality $Ti_3C_2T_x$ films with a thickness as low as 8.6 nm on flat substrates, such as glass, silicon (Si) wafers, and aluminum (Al) plates. After the fabrication, we measured their infrared (IR) optical properties experimentally, including the broadband complex permittivity of various $Ti_3C_2T_x$ films over the wavelength region of 0.3-20 μm and their IR properties' limitation of reflectance and absorptance/emittance in 2.5-16.7 μm. As a result, a 568-nm-thick $Ti_3C_2T_x$ film shows large and negative real part $\varepsilon_1(\omega)$, as well as large $\varepsilon_2(\omega)$ of complex permittivity in the IR region, yielding metal-like properties with the maximum IR reflection of 88%, and low emittance of 12%. For $Ti_3C_2T_x$ films of different thickness, the variation of their IR reflectance is attributed to not only the film thickness, but also their variation of complex permittivity. This phenomenon is because thinner $Ti_3C_2T_x$ films yield lager interlayer space induced in the fabrication process, leading to larger total dipole moments, and finally larger values of both real and imaginary parts according to the first-principles molecular dynamics model. Such metallic properties are favorable to produce high IR reflectance for ultrathin $Ti_3C_2T_x$ films, (e.g., 61.6% of the IR reflectance for a 28.8-nm-thick $Ti_3C_2T_x$ film). By vacuum-assisted filtration of $Ti_3C_2T_x$ films on IR-transparent substrates, i.e., nanoporous polyethylene (nanoPE) films, we also investigated the IR absorptance/emittance of ultrathin $Ti_3C_2T_x$ films. Their IR absorptance increases with thickness until the maximum of 37.5% for a 15-nm-thick $Ti_3C_2T_x$ film and then decreases until thickness-independent 12% of thick films. This phenomenon can be clarified that more inelastic collisions happen in restricted ultrathin films and collided electrons can absorb more IR photons to excite themselves. Our findings provide an avenue for the preparation of large-scale ultrathin two-dimensional (2D) films within optical quality to further study various characteristics. Due to superb electrical and metal-like optical properties, this work will stimulate more research about $Ti_3C_2T_x$ for multiple energy-saving applications.

**Methods**

**3.1 Optical spectrum**

Ultrathin $Ti_3C_2T_x$ films have high transmittance in the visible region. With the increase of thickness, more solar light would be absorbed, and most infrared (IR) light was reflected by $Ti_3C_2T_x$ films. In this work, a UV-Visible-NIR spectrometer (Lambda 950, Perkin Elmer) was utilized to measure the spectrum of $Ti_3C_2T_x$ films transferred on glass over the wavelengths of 0.3-2.5 μm. The glass substrate can transmit around 92% of solar light in 0.3-2.5 μm without absorption. Therefore, the intrinsic transmittance of $Ti_3C_2T_x$ films equals to the ratio of total transmittance of $Ti_3C_2T_x$ on glass to the transmittance of glass, defined as: $T_{\text{MXene}} = \frac{T_{\text{MXene+glass}}}{T_{\text{glass}}}$. In the mid-IR region varying from 2.5 to 16.7 μm, the reflectance of $Ti_3C_2T_x$ films on filter membranes was tested by a gold integrating sphere (PIKE Technologies) using the Fourier Transform Infrared Spectroscopy (FTIR, Vertex 70, 275 Bruker). Commercial polyvinylidene difluoride (PVDF) membranes were applied to support $Ti_3C_2T_x$ films for the measurement of mid-IR spectrum due to their negligible IR reflectance of less than 2%.

### 3.2 Ellipsometric spectrometry

The refractive index measurement in the NIR-MIR region (1.5-20 μm) was performed via an ellipsometer (IR-VASE MARK II, J.A. Woollam Co., Inc.). The $Ti_3C_2T_x$ film with a determined thickness was vacuum-assisted filtrated on polymer membranes and then transferred on pre-cleaned glass substrates. The thickness of $Ti_3C_2T_x$ films was measured by AFM or SEM. In the ellipsometric spectrometry, the integral sample of $Ti_3C_2T_x$ films on glass was illuminated via a beam of polarized light without destruction, and the measurement was performed over a spectral range from 300 to 7000 cm$^{-1}$ with a step size of 16 cm$^{-1}$. The optical constants of samples were determined using WVASE software [37]. The modeling procedure in this software is initiated through the construction of a layered structure, and substrates (i.e., glass) in this study was simplified as a semi-infinite slab. It should be noted that the optical properties of all samples were eventually allowed to vary in certain ranges because of the variable film thickness determined by surface roughness, as well as the roughness of various substrates. To obtain the best approximation to the intrinsic dielectric functions ($\tilde{\varepsilon}$) of films, the $\tilde{\varepsilon}$ of glass was constructed using a multi-oscillator model first. Then, we used B-splines to describe $\tilde{\varepsilon}$ of our $Ti_3C_2T_x$ films. B-splines can effectively describe complex structures in absorptance spectra while simultaneously preserving Kramers-Kronig consistency [26].

### 3.3 Sheet resistance

Samples for the sheet resistance test were fabricated directly on porous filter membranes via vacuum-assisted filtration. The sheet resistance of various $Ti_3C_2T_x$ films was measured by four-point probe method over 10 cycles at room temperature (Ecopia HMS-5500). The corresponding electrical properties, i.e., carrier concentration, mobility, and sheet resistance, were defined as the average value of ten sets of data.

**3.4 X-ray diffraction measurement**

The interlayer space between $Ti_3C_2T_x$ flakes was determined by X-ray diffractometer (X'pert Pro, PANalytical) with the Cu Kα wavelength of around 1.54 Å. The scan angle ranges from 5 to 20° with a slow scan rate of 0.013 degrees per second. To ensure the consistency between interlayer space and optical properties, the same set of samples on glass used for the ellipsometric spectrometry measurement were also employed for the XRD test. Glass were processed vis ultrasonic cleaning in three types of solutions of ethanol alcohol, acetone, and DI water successively.

**Acknowledgments**

This work was finically supported from the Hong Kong General Research Fund (Grant Nos. 16206020) and the Center on Smart Sensors and Environmental Technologies (Grant No. IOPCF21EG01) in the Hong Kong University of Science and Technology. The authors are also thankful for partly supported by the Project of Hetao Shenzhen- Hong Kong Science and Technology Innovation Cooperation Zone (HZQB-KCZYB-2020083).